\documentclass[preprint,12pt,times,authoryear]{elsarticle}
\usepackage{amssymb, amsmath}
\usepackage{subcaption}
\usepackage{rotating}
\usepackage{pdflscape}
\usepackage{tabularx}
\usepackage{float}
\usepackage{booktabs}
\usepackage{changepage}
\newlength{\extralength}
\newlength{\fulllength}
\begin{document}

\title{Export proceeds repatriation policies: A shield against exchange rate volatility in emerging markets?}

\author[feb,bi]{Sondang Panggabean\corref{cor1}}
\ead{smarshinta@bi.go.id}

\author[feb]{Mahjus Ekananda}
\ead{machjus.ekananda@ui.ac.id}

\author[feb]{Beta Yulianita Gitaharie}
\ead{beta.yulianita@ui.ac.id}

\author[bi]{Leslie Djuranovik}
\ead{leslie@bi.go.id}
\cortext[cor1]{Corresponding author.}
\affiliation[feb]{organization={Faculty of Economics and Business, Universitas Indonesia},
            city={Depok},
            country={Indonesia}}
\affiliation[bi]{organization={Bank Indonesia},
            city={Jakarta},
            country={Indonesia}}

\begin{abstract}
We examine the impact of mandatory export proceeds repatriation on exchange rate stability in three emerging markets, Iran, Sri Lanka, and Turkey, using the Generalized Synthetic Control framework. By modeling exchange rate stochastic volatility as our outcome of interest and controlling for interest rate differentials, exchange rate regime shifts, and inflation rate gaps, we address both unobserved time‐varying confounders and heterogeneous treatment effects. Our estimates reveal no statistically significant impact of repatriation mandates on exchange‐rate volatility across the three countries. We also find that we cannot reject the possibility of a non-zero impact. These results remain robust to an extensive range of sensitivity analyses, including alternative covariate specifications and placebo tests, thereby confirming their reliability.
\end{abstract}

\maketitle

\section{Introduction}
\label{intro}
Exchange rate volatility is a critical concern, especially for emerging economies, where fluctuations in currency values can significantly impact the performance of an economy \citep{Krol2014}. Greater exchange rate volatility slows productivity and gross domestic product growth \citep{Aghion2009, Bush2021}. There is also evidence that higher exchange rate volatility affects the unemployment rate of a country \citep{Feldmann2011}, raises transaction risk related to international trade \citep{Baum2010}, and reduces consumption and investment \citep{Grier2006}. One of the key policy measures employed to stabilize exchange rate movements is the export proceeds repatriation (EPR) policy, which mandates that exporters bring their foreign exchange earnings back into the exporters' country within a specified period. 

We apply the generalized synthetic control (GSC) estimator of \citep{Xu2017} to rigorously quantify the effectiveness of mandatory export proceeds repatriation in stabilizing exchange rate volatility. By integrating interactive fixed effects with a synthetic control framework, the GSC flexibly adjusts for both unobserved time‑varying confounders and heterogeneous treatment effects, delivering a credible counterfactual analysis path for each treated country. Because GSCs require single, non‐reversible treatments and sufficiently long pretreatment periods, our analysis focuses on Iran, Sri Lanka, and Turkey, three emerging economies that implemented continuous repatriation rules during our observation period, which is March 2008 to December 2021.

Iran began implementing an EPR policy in April 2018 with no requirement to convert export proceeds into local currency. The repatriation rules differ depending on the amount of funds received by exporters. The difference covers the maximum percentage of export proceeds that may be converted into Iranian rial based on an exporter's foreign transaction, while the rest should be used for imports. In January 2014, Sri Lanka implemented an EPR policy requiring the conversion of the proceeds into Sri Lankan rupees. However, Sri Lankan exporters are still allowed to retain the proceeds abroad, provided that the fund is not used to buy properties or other capital assets abroad. On the other hand, Turkey has a long history of implementation. Turkey has implemented various iterations of its EPR policy over the years. The policy was initially introduced in the 1980s as part of broader economic liberalization measures aimed at integrating Turkey into global financial markets. However, as the country experienced recurrent currency crises, the policy underwent several modifications. By March 2008, the policy had been retracted, and there was no obligation to repatriate export proceeds. In September 2018, amid heightened exchange rate volatility and increasing external debt pressures, the government reinstated strict repatriation requirements, mandating that at least 80\% of export earnings be brought back to Turkey within 180 days.

These regulations were introduced to channel exporters’ foreign-currency earnings back into the domestic market, limit speculative depreciation, and support macroeconomic stability. However, whether they succeed in reducing exchange-rate volatility, remains unclear.  The outcomes are confounded by concurrent forces such as geopolitical shocks, COVID-19 pandemic, episodes of global risk aversion, and persistent domestic inflation. Therefore, the specific contribution of EPR rules to currency stability has received little systematic scrutiny. This study addresses that gap; to the best of our knowledge, it is among the first to provide a rigorous, policy-focused evaluation of EPR effects on exchange-rate volatility in an emerging-market setting.

We analyze monthly data for the three adopters and ask whether the EPR mandated dampened volatility or was overshadowed by broader macrofinancial conditions. Our empirical design tests the policy’s impact while explicitly controlling for interest-rate differentials, the prevailing exchange-rate regimes, and inflation gaps. By disentangling these channels, this study clarifies how capital flow regulations interact with domestic and external factors to shape currency dynamics. The findings inform policy debates on capital-flow management, highlighting the circumstances under which EPR rules can, or cannot, enhance exchange-rate stability in emerging economies. These findings could help policy makers formulate better policies regarding EPR that would have a greater impact on exchange rate stability. 

The remainder of the paper is organized as follows. Section 2 summarizes the key literature on export proceeds and exchange rate volatility. Section 3 introduces the data and describes our empirical strategy. Section 4 presents the results and the empirical analysis. Section 5 discusses the robustness tests. Finally, Section 6 concludes the paper. 

\section{Literature Review on Export Proceeds and Exchange Rate Volatility} 
By 2021, 83 of 156 \footnote{Data source from The Annual Report on Exchange Arrangements and Exchange Restrictions by the IMF} emerging countries had adopted EPR policies, yet research on export proceeds remains scarce. Early studies emphasized stabilizing export earnings amid commodity price and quantity fluctuations \citep{Wallich1961, Powell1959, Massell1964, Macbean1962}. Later work examined export proceeds as part of the trade balance \citep{Guillaumont1980} and as a factor influencing foreign direct investment \citep{Asiedu2004, Prayoga2024}. This paper contributes to the limited body of literature by investigating how EPR policies affect exchange rate volatility. 

The majority of exchange rate literature on developing economies focuses on how exchange rates influence growth. It also examines how institutional factors, such as international trade and financial openness or exchange rate regimes, shape exchange rate fluctuations. A consistent finding is the negative impact of exchange rate volatility on long-term economic performance \citep{Devereux2002, Choudhri2006, Grier2006, Lee-lee2007, Aghion2009, Krol2014, Feldmann2011, Bush2021, Aysun2024, Sukmawati2018}. In contrast, this study examines EPR policies as drivers of exchange rate volatility.

By design, EPR requirements direct additional export earnings back into the domestic financial system, thereby expanding the supply of foreign exchange \citep{Calvo1996} regardless of whether those receipts are subsequently converted into local currency. This transmission highlights the relationship between foreign exchange capital flows and exchange rate volatility, which is our outcome of interest. Some research indicates different volatility determinants in developing economies from determinants in developed countries \citep{Grossmann2014a}. For example, \citep{Flood1999} argue that macroeconomic factors are less critical in developing economies, whereas \citep{Rafi2018} and \citep{Caporale2017} emphasize the importance of foreign capital flows. Furthermore, \citep{Gabaix2015} develop a theoretical model that also supports the importance of foreign capital flows, influenced by relative levels of interest rates, in determining exchange rate volatility. The model is also disconnected from traditional macroeconomic fundamentals. \citep{Brunnermeier2008} and \citep{Farhi2016} also describe exchange rate dynamics through changes in the interest rate differentials (IRD) that lead to movements in capital flows. 

Furthermore, the intensity of exchange rate volatility is a direct consequence of exchange rate regimes \citep{Flood1999, MacDonald2007, Sarno2003} where pegged exchange rate regimes will produce lower exchange rates than floating regimes (\citep{Ghosh2002}, \citep{Levy-Yeyati2003}, inter alia). However, soft-pegged and managed floating regimes are still vulnerable to the volatility of exchange rates. Countries that adopted EPR policies also adopted soft-pegged and managed floating regimes as listed in Table \ref{regime}. 

Inflation differentials vis-\`a-vis the United States are also important in determining exchange rate volatility via purchasing power parity (PPP).  In contrast to most studies that cite the impact of exchange rate volatility on inflation, \citep{Farhi2016} develop a theoretical model that proves the influence of inflation on exchange rate volatility during rare disasters. The comovement between inflation and exchange rates is also empirically shown by \citep{Sen2020}. Moreover, because PPP prevails over time, a burst of inflation eventually weakens the currency, even though it might initially strengthen; in the classic overshooting model, where the central bank fixes the money-supply growth rate, an inflation surprise is therefore expected to end up pushing the exchange rate down or in words, a depreciation \citep{Clarida2008}. \citep{Ekananda2021} show that the effect of global inflation on domestic inflation may differ at different threshold levels, causing the effect of inflation on exchange rate volatility to also vary. In short, inflation (or similarly inflation differential) affects exchange rate volatility. \citep{Clarida2002} also theoretically show that in a two-country Nash equilibrium under central bank discretion, inflation does affect exchange rate volatility. Similar results from empirical studies are also given by \citep{Goldberg2005} and \citep{Faust2007}. 

Furthermore, \citep{Orlov2006} and \citep{Orlov2009} show that studying volatility via a conventional time-domain approach (i.e., variance or standard deviation) may lead to spurious results if the effect on various components of volatility is not uniform. Among others, \citep{Danielsson1998}, \citep{Iseringhausen2020}, and \citep{Kim1998} show that given the choice of statistic volatility models and generalized autoregressive conditional heteroskedasticity (GARCH) models, stochastic volatility models are preferable, given their efficiency and better forecasting power. These properties are essential for both regulators and market players.

Prior evidence suggests that capital flow restrictions seldom dampen exchange rate fluctuations. \citep{Glick2005} and \citep{Edwards2009} find little support for a volatility‑reducing effect of capital controls, while \citep{Panggabean2024} report that Indonesia’s repatriated export proceeds leave short‑ and medium‑term exchange rate volatility unchanged. Guided by these findings, we propose the following hypothesis for the present study:

\textbf{$H_0$:} Export‑proceeds repatriation (EPR) mandates in Iran, Sri Lanka, and Turkey have no measurable effect on monthly exchange‑rate volatility.
The empirical analysis that follows tests this null against the alternative that EPR policies either increase or decrease volatility in the treated economies.

\section{Data and empirical methodology}
\label{method}

\subsection{Data}
\label{data}
\begin{figure}[h!]
	\includegraphics[width=\textwidth]{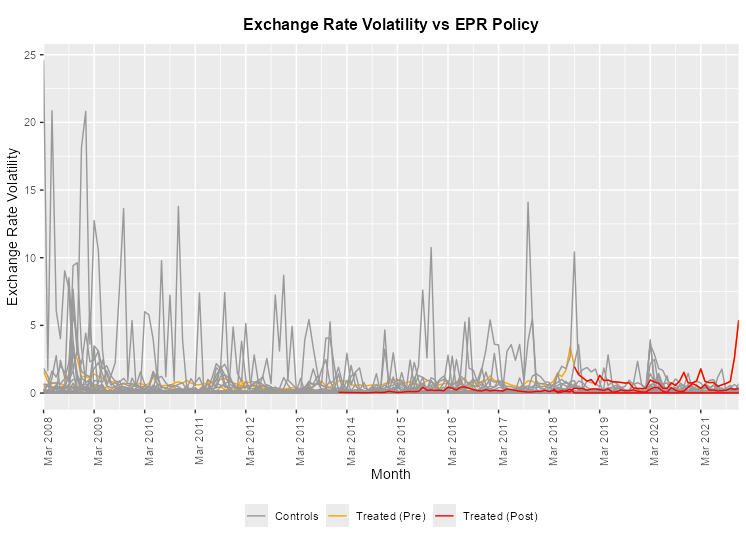}
	\caption{Exchange rate volatility of all sample countries\label{fig:volatility}}	
\end{figure}

\begin{figure}[h!]
	\centering
	\includegraphics[width=\textwidth]{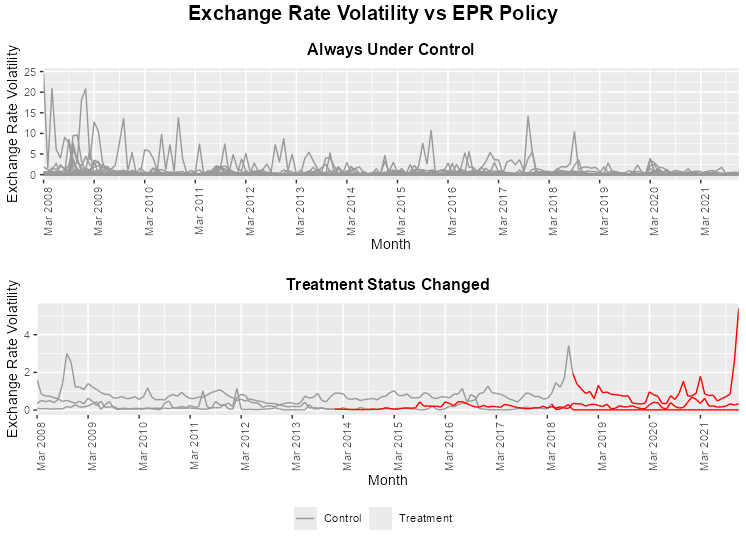}
	\caption{Exchange rate volatility of control and treated countries\label{fig:volatility2}}	
\end{figure}

We retrieve data regarding the timing of the implementation of EPR policies in Iran, Sri Lanka, and Turkey from The Annual Report on Exchange Rate Arrangements and Exchange Restrictions (AREAER) database of the International Monetary Fund (IMF). Iran, Sri Lanka, and Turkey present an ideal case for analysis, given their consistent implementation of the policy without reversals since March 2008 to December 2021 and their adherence to a non-hard-pegged exchange rate regime as classified by the IMF during our observation period. We follow the general guidance proposed by \citep{Abadie2015} to choose our treated and control units. First, the treated units should not experience large idiosyncratic shocks to exchange rate volatility (outcome of interest) as a result of the reform implementation of EPR policies (our treatment). We observe the absence of exchange rate volatility shocks caused by EPR policies implementation in Figure~\ref{fig:volatility} and the second plot in Figure~\ref{fig:volatility2}. Second, the analysis requires a large number of pretreatment and posttreatment observations. Table~\ref{observation} summarizes the number of observations of the treated units.

\noindent{\footnotesize{Note: Exchange rate volatility is calculated by implementing the stochastic volatility model}}

\noindent{\footnotesize{Note: Exchange rate volatility is calculated by implementing the stochastic volatility model}}

\begin{table}[h!]
\caption{Number of monthly observations of treated countries\label{observation}}
\begin{tabularx}{\textwidth}{XXX}
\toprule
\textbf{Treated Country} & \textbf{Number of Pretreatment Observations} & \textbf{Number of Posttreatment Observations}\\
\midrule
Iran                                                                                     & 121                                                                                      & 45                                                                                        \\
Sri Lanka                                                                                & 70                                                                                       & 96                                                                                        \\ 
Turkey                                                                                   & 126                                                                                      & 40                                                                                        \\
\bottomrule
\end{tabularx}
\end{table}

Third, the control units must have never received treatment, which, in our case, are the countries that never implemented EPR policies during our observation period. Finally, the control units should be structurally similar to the treated units. It is important to avoid bias resulting from comparing countries with different characteristics. Hence, we select only countries categorized as emerging countries by the IMF, which is consistent with our choice of data source for de facto exchange rate regimes \footnote{\citep{Anderson2009} gives the complete list and definition for de facto exchange rate regimes} and countries implementing EPR policies. The countries that comply with the third and fourth requirements are twenty-four developing countries that never applied the export proceeds repatriation policy from March 2008 to December 2021. The countries have been additionally filtered by de facto exchange rate regimes, as defined by the IMF \citep{Anderson2009}, which excludes those countries that apply hard-pegged exchange rate regimes. We avoid countries that adopt hard-pegged exchange rate regimes since it is not relevant to analyze exchange rate volatility for such countries \citep{Flood1999, MacDonald2007, Sarno2003}. The list of treated and control units, along with their exchange rate regimes in alphabetical order, is given in Table~\ref{regime}.

\begin{table}[h!] 
\caption{Exchange rate regimes of treated and control countries}
\label{regime}
\begin{tabularx}{\textwidth}{cXcc}
\toprule
\textbf{No.} 	& \textbf{Country} 	&\textbf{Exchange Rate Regimes}\textsuperscript{1} &\textbf{Status} 		\\ 
\midrule
1            		& Armenia                                    		& 	1, 2			& Control							 \\ 
2            		& Bahrain                                    		& 	1				& Control						 \\ 
3            		& Bolivia                                   		& 	1				& Control						 \\ 
4            		& Botswana                                 		& 	1				& Control						 \\ 
5            		& Gambia                                     		& 	0, 1, 2			& Control							 \\ 
6            		& Georgia                                   		& 	0, 1, 2			& Control							 \\ 
7				& Iran								&	0, 1				& Treated								 \\ 
8            		& Jamaica                                    		& 	1, 2				& Control						 \\ 
9            		& Jordan                                    		& 	1				& Control						 \\ 
10            		& Kenya                                  			& 	0, 1, 2		& Control								 \\ 
11            		& Kuwait                                    		& 	0, 1				& Control						 \\ 
12           		& Kyrgyzstan                                   		& 	0, 1, 2		& Control								 \\ 
13            		& Mexico                                    		& 	2				& Control						 \\ 
14            		& Mongolia                                    		& 	0, 1, 2		& Control								 \\ 
15            		& Oman                                   		& 	1				& Control						 \\ 
16            		& Paraguay                                     		& 	0, 1, 2		& Control								 \\ 
17            		& Peru                                    			& 	1, 2			& Control							 \\ 
18            		& Philippines                                    		& 	1, 2			& Control							 \\ 
19            		& Qatar                                    		& 	1				& Control						 \\ 
20            		& Saudi Arabia                                   	& 	1				& Control						 \\ 
21				& Sri Lanka							&	1, 2				& Treated						\\ 
22            		& Trinidad and Tobago                              	& 	1			& Control							 \\ 
23				& Turkey							&	2				& Treated						\\ 
24            		& Uganda                                   		& 	2				& Control						 \\ 
25            		& United Arab Emirates                            	& 	1			& Control							 \\ 
26            		& Uruguay                                    		& 	2			& Control							 \\ 
27            		& Zambia                                   		& 	1, 2				& Control						 \\ 
\bottomrule
\end{tabularx}
\noindent{\footnotesize{\textsuperscript{1}1=soft-pegged (2009 de facto system: conventional pegged arrangement, stabilized arrangement, pegged exchange rate within horizontal bands, crawling peg, crawl-like arrangement); 2=floating (2009 de facto system: floating, free floating); 0=other managed arrangements}}
\end{table}

To examine the impact of EPR policies on exchange rate volatility, this paper utilizes the GSC with four main covariates, namely interest rate differential (IRD), inflation differential, and exchange rate regime in the form of a categorical covariate.  Since GSC does not allow treatment reversal, we began our monthly observation in March 2008 until December 2021, purposefully avoiding treatment reversal for Turkey while allowing sufficient time for a pretreatment period. 

Exchange rate volatility is calculated via the stochastic volatility model following \citep{Kim1998} with daily exchange rates retrieved from Refinitiv Eikon from 1 March 2008 to 31 December 2021. The daily exchange rate volatility is then converted into monthly volatility by taking the square root of the average of the sum of the squared daily stochastic volatility for each month. Interest rate differentials are calculated as the difference between the effective federal funds rate (EFFR) \footnote{https://fred.stlouisfed.org/series/EFFR} and a country's short-term interest rates, both compounded into monthly rates. The interest rates are retrieved from several sources, depending on data availability, with preferences given to interbank short-term rates, as listed in Table~\ref{ird} in the Appendix. Moreover, monthly inflation data are retrieved from the IMF database. 

\subsection{Methodology}
\label{methodology}
\subsubsection{Stochastic Volatility}
The exchange rate volatility is estimated following the stochastic volatility model parameters. The model for regularly spaced data is
\begin{equation}\label{svstandard}
	y_t=\beta e^{\frac{h_t}{2}} \epsilon_t 
\end{equation}
where $y_t$ is the mean-corrected exchange rate return with $t=1, ..., T$ and $\epsilon_t \sim N(0,1)$. 

The conditional variance of $y_t$ given $h_t$ is $Var(y_t|h_t) = e^{h_t}$,
implying that the conditional variance is time-varying. We assume that $h_t$, the log-volatility of the exchange rate return $y_t$ at time $t$, follows 
\begin{equation}\label{logvolsv}
h_{t+1} = \mu + \phi (h_t - \mu) + \sigma_{\eta} \eta_t
\end{equation}
where $\phi$ can be thought as the persistence in the volatility, $\sigma_{\eta}$ is the volatility of the log-volatility, and $\eta_t \sim N(0,1)$ and $h_1 \sim N \left( \mu, \frac{\sigma^2}{1-\phi^2} \right)$. For identifiability reasons, following \citep{Kim1998}, $\beta$ will be set to one when we estimate the model. Finally, $\epsilon_t$ and $\eta_t$ are uncorrelated standard normal noise shocks.  

Equation \ref{svstandard} can be rewritten as
\begin{equation}\label{logsv}
	\log y_t^2 = \mu + h_t + \log \epsilon_t^2.
\end{equation}
Thus, the estimation of stochastic volatility models requires the estimation of the parameters 
\begin{equation}\label{parametersv}
	\theta=\{ \mu, \phi, \sigma_{\eta} \}.
\end{equation}
We use an auxiliary mixture sampler following \citep{Kim1998} to estimate them.

Since volatility is a latent variable, we need to compute $h_t|Y_t,\theta$ for each value of $Y_t=(y_1, \dots , y_t)$ by particle filtering. We employ the particle filtering algorithm following \citep{Kim1998} and \citep{Pitt1999}.  We modify MATLAB codes from \citep{Chan2019} using $20,000$ iterations and discard the first 1,000 iterations for convergence.

\subsubsection{Generalized Synthetic Control}
\begin{figure}[h!]
	\includegraphics[width=\textwidth]{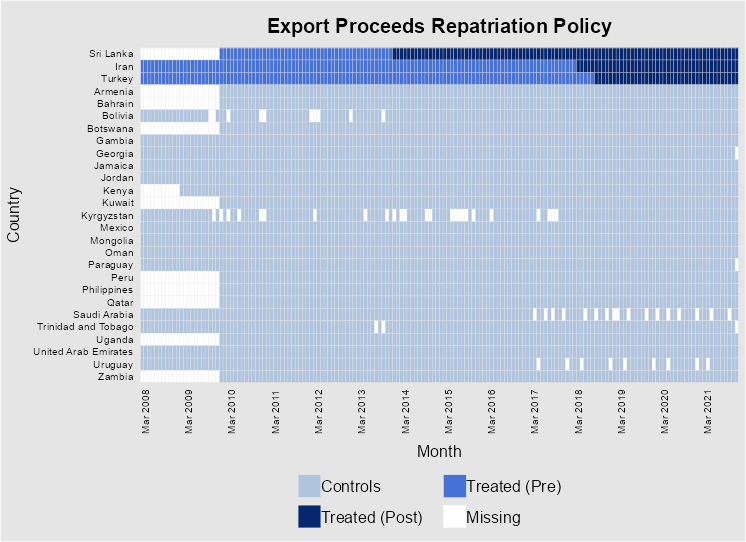}
	\centering
	\caption{Implementation of export proceeds repatriation policies in treated and control countries}
	\label{fig:panelview}
\end{figure}

Our study applies GSC following \citep{Xu2017}, which relaxes the parallel trends assumption commonly required in Difference-in-Differences (DID) (\citep{Card1990, Card1994, Angrist2009}, inter alia) and simultaneously unifies the synthetic control method \citep{Abadie2010, Abadie2015}. In a DID design, the parallel trends assumption requires that, had the treated units never received the intervention, their outcome of interest path would have shared the same baseline slope as that observed for the control group; hence any post-intervention divergence in slopes is attributed to the ATT. Although Figure~\ref{fig:panelview} suggests that a conventional DID strategy could be applied to our three treated countries, the impact of export proceeds repatriation policy on exchange rate volatility is subject to unobserved, time varying confounders. These confounders violate the parallel trends assumption, making a DID specification including the staggered DID design proposed by \citep{Wing2024}, inter alia, unsuitable for this analysis. As our treated units are more than one, using the synthetic control method proposed by \citep{Abadie2010} is also not feasible.

In order to handle the bias due to the existence of unobserved time-varying confounders, earlier research suggested the use of factor-augmented estimators \citep{Bai2009, Gobillon2016, Liu2024, Xu2017}. The factor-augmented estimator we employ is the interactive fixed effect (IFE) model \citep{Bai2009}. The IFE model employs the interaction between unit-specific intercepts (factor loadings) with time-varying coefficients (latent factors).  \citep{Gobillon2016} show the superiority of the IFE model over the synthetic control method in DID settings when unit-specific intercepts (factor loadings) do not share a common support. The GSC incorporates the IFE into the synthetic control model.  

The GSC estimator is specifically designed to remain stable in the presence of limited observations \citep{Xu2017}. The robustness makes it an appropriate choice for our dataset, which contains just three treated countries and twenty‑four control countries, each observed over 166 monthly periods. Furthermore, the GSC does not discard any observations from the control group; thus, it uses more information from the control group, resulting in a more efficient procedure than the synthetic control method does. A cross-validation scheme is also embedded in the GSC model. This scheme automatically selects the number of factors of the IFE model with high probability, thereby reducing the risk of overfitting.

The framework of the GSC model relies on two main assumptions. First, it requires the treated and control units to be affected by the same set of factors, while the number of factors is fixed for the entire observed period. Thus, it cannot capture unobserved confounders that are independent across units. Second, it assumes that the error term of any unit at any time is independent of the treatment, observed covariates, and unobserved cross-sectional and temporal heterogeneities of all units (including itself) at all periods of observation \citep{Xu2017, Liu2024}. 

In our study, the outcome of interest is exchange rate volatility, the observation unit is the country (treated and controlled), and the time is the month of observation. The covariates are IRD, inflation differential, and exchange rate regime. We do not include interactions with time and unit (country) since two-way fixed effects in our design absorb any common time-only shocks across units and unit-only shocks across time. Finally, our treatment is the implementation of EPR policies. 

Let $Y_{it}$ be our outcome of interest of unit $i$ at time $t$, $N$ is the total number of units with $N=N_{treat} + N_{ctrl}$ where $N_{treat}$ is the number of treated unit and $N_{ctrl}$ is the number of control units. The control units are subscripted from $1$ to $N_{ctrl}$ and the treated units from $N_{ctrl}+1$ to $N$. The number of periods of observations is $T$ with $T_{0, i}$ as the number of pretreatment periods of treated unit $i$ while $T_0$ is the period of observation at which pretreatment ends. Hence, we denote the time at which unit $i$ is first treated as $T_{(0,i) + 1}$ and control units are never treated at all time $t$. A summary of our observations is given in Table~\ref{obs}. In addition, let $\mathcal{T}$ denotes the set of units in the treated group and $\mathcal{C}$ denotes the set of units in the control group.

\begin{table}[h!]
\caption{Summary of observation data}
 \label{obs}
\begin{tabularx}{\textwidth}{Xcc}
\toprule
\textbf{Unit}  & \textbf{$T_{0,i}$} & \textbf{N} \\ 
\midrule
Treated unit 1                       & 45                 & $N_{treat}=1$          \\ 
Treated unit 2                       & 96                 & $N_{treat}=1$          \\ 
Treated unit 3				 & 40                 & $N_{treat}=1$          \\ 
Control Unit                         & 166                & $N_{ctrl}=24$         \\ 
\bottomrule
\end{tabularx}
\end{table}

The functional form of our design, following \citep{Xu2017}, is

\begin{equation} \label{functional}
	Y_{it} = \delta_{it} D_{it} + X_{it}' \beta + \lambda_i'f_t + \epsilon_{it},
\end{equation}
where $\delta_{it}$ is the heterogeneous treatment effect on unit $i$ at time $t$; $D_{it}$ is the treatment indicator, which equals $1$ if unit $i$ is treated at time $t$ and equals $0$ otherwise; $X_{it}$ is a $(3 \times 1)$ vector of observed covariates; $\beta$ is a $(3 \times 1)$ vector of unknown parameters which, in this study, is assumed to be constant across space and time mainly for faster computation \footnote{as noted by \citep{Xu2017}, this limitation can be further addressed by using random coefficient models in Bayesian multi-level analysis; however, in this study constant $\beta$ is fairly acceptable since any variation across unit and time are already absorbed by the fixed effects};  $\lambda_i$ is an $(r \times 1)$ vector of unknown factor loadings; $f_t$ is an $(r \times 1)$ vector of unobserved common factors with $r$ is the number of factors; and $\epsilon_{it}$ is an unobserved idiosyncratic shock for unit $i$ and time $t$ with mean zero.

Our main quantity of interest is the average treatment effect on the treated (ATT) at time $t$ when $t > T_0$, given by
 
\begin{equation} \label{eq:att}
	ATT_{t,t>T_0}=\frac{1}{N_{treat}} \sum_{i \in \mathcal{T}} [\mathbb{E} [Y_{it}(1)|D_i=1] -\mathbb{E} [Y_{it}(0)|D_i=1]] =\frac{1}{N_{treat}} \sum_{i \in \mathcal{T}} \alpha_{it},
\end{equation}
where $Y_{it}(1)$ and $Y_{it}(0)$ are potential outcome if unit $i$ is treated at time $t$ and  potential outcome if unit $i$ is not treated at time $t$,  respectively. As in \citep{Abadie2010}, we do not incorporate the uncertainty of the treatment effects $\alpha_{it}$ since we are making inferences about the ATT in the sample, not the ATT of the population. Finally, the model is implemented by using \textit{gsynth} R package \citep{Xu2017}.
\section{Empirical results and discussions}
\subsection{Results}\label{result}
Our cross‐validation procedure indicates that three factors ($r = 3$) minimize the mean squared prediction error (MSPE), the outcomes of which are reported in Table~\ref{factor}. In reference to Equation \ref{functional}, $\lambda_i$ is then a $(3 \times 1)$ vector of unknown factor loadings and $f_t$ is a $(3 \times 1)$ vector of unobserved common factors.

\begin{table}[h!]
\caption{Cross-validation outcomes}
 \label{factor}
\begin{tabularx}{\textwidth}{ccccc}
\toprule
\textbf{r} &\textbf{sigma2} & \textbf{IC} & \textbf{PC} & \textbf{MSPE} \\
\midrule 
0                                & 0.96516                            & 0.38998                          & 0.91520                          & 0.13973
 \\ 
1                                & 0.21216                              &  -0.71061                         & 0.27316                          & 0.09043
 \\ 
2                                & 0.12944                              & -0.79481                         & 0.21065                          & 0.08381
 \\ 
3*                                & 0.10236                              & -0.62404                         & 0.20142                         & 0.08024
 \\ 
4                                & 0.08534                              & -0.40479                         & 0.19703                          & 0.08485
 \\ 
\bottomrule
\end{tabularx}
\noindent{\footnotesize{ Note: r is the number of unobserved factors; sigma2 is the estimated variance of the error term; IC is the Bayesian Information Criterion; PC is the penalty criterion.}}
\end{table}

\begin{figure}[h!]
	\includegraphics[width=\textwidth]{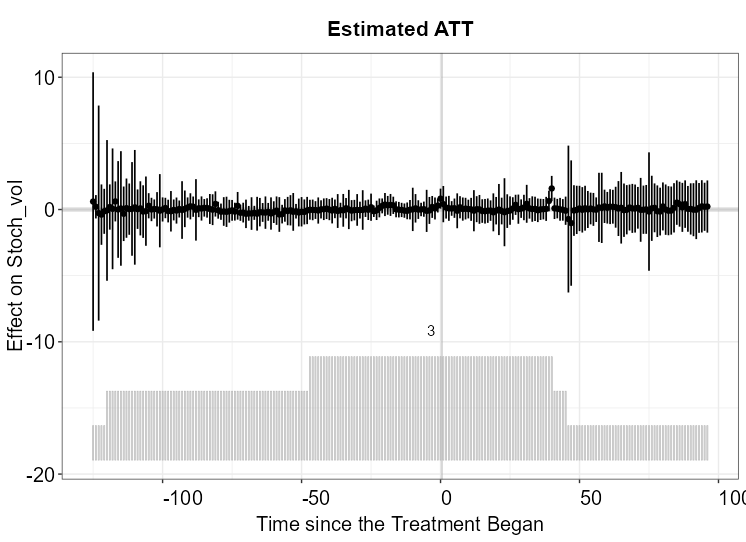}
	\caption{Estimated average treatment effect of the treated}
	\label{fig:att}
\end{figure}

 Table~\ref{att} and Figure~\ref{fig:att} present the estimated ATT and its temporal evolution relative to the start of treatment. The estimated p‑value of $0.65541$,  exceeds the conventional  5\% threshold for statistical significance. Accordingly, we fail to reject the null hypothesis that the EPR policy has no effect on exchange rate volatility. Adopting a conservative stance that adheres to the 5\% significance level (see \citep{Benjamin2019}), we infer that the EPR policy does not exert a statistically robust influence on exchange‑rate volatility across the three treated countries. This inference is reinforced by the estimated ATTs, whose confidence intervals almost uniformly span zero, underscoring both the statistical imprecision and the lack of an economically meaningful effect. In sum, the available evidence remains insufficient to substantiate any substantive influence of the EPR policy on exchange‑rate volatility.

\begin{table}[h!]
\caption{Estimated average treatment effect on the treated}
 \label{att}
\begin{tabularx}{\textwidth}{ccccc}
\toprule
\textbf{Average ATT} & \textbf{Standard Error} & \textbf{Lower CI} & \textbf{Upper CI} & \textbf{p-value} \\ 
\midrule
0.06939                      & 0.15550                          & -0.23538              & 0.37417                & 0.65541       \\ 
\bottomrule
\end{tabularx}
\noindent{\footnotesize{Note: CI = 95\% confidence interval}}
\end{table}

\begin{figure}[h!]
	\includegraphics[width=\textwidth]{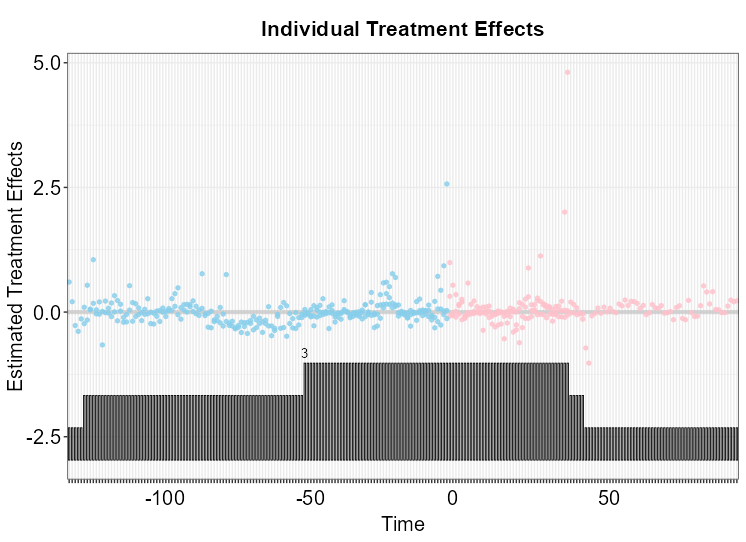}
	\caption{Estimated individual average treatment effect on the treated}
	\label{fig:individual}
\end{figure}

Figure~\ref{fig:individual} shows the estimated individual treatment effects for every treated unit and period, centered on the policy-adoption date (t = 0). Pretreatment estimates oscillate tightly around zero, with most points lying within ±0.5 standardized units and no discernible drift, indicating a good pre-fit and no anticipatory effects. After adoption, the cloud of points remains centered near zero and retains a similar vertical spread; only a handful of positive outliers peaking at approximately +5 appear several periods after the shock. These idiosyncratic spikes are sparse and do not shift the modal treatment effect, which remains statistically indistinguishable from zero. The accompanying histogram shows that most observations cluster between t = –50 and t = +30; within this dense window, the dot cloud still straddles the zero line, implying that a larger sample mass does not reveal a latent effect concealed by sampling noise. The plot suggests that the export-proceeds repatriation policy did not produce a systematic or economically meaningful change in exchange-rate volatility across treated units; any detectable impact is confined to a few outlying episodes rather than the typical experience. Thus, at the individual level, the policy’s impact is neither widespread nor uniformly strong.

With respect to Equation~\ref{functional}, Table~\ref{beta} presents our estimates of the covariate coefficients, $\beta$. By observing the covariates and the p-values, we conclude that both IRD and the exchange rate regime are statistically significant positive predictors of exchange rate volatility with robust standard errors. In contrast, the inflation differential is not statistically significant, as its coefficient is indistinguishable from zero, from our observation of the p-value and the confidence interval that crosses zero. These results suggest that for our sample countries, inflation differentials do not play an important role in determining exchange rate volatility.

\begin{table}[h!]
\caption{Estimated $\beta$}
 \label{beta}
\begin{tabularx}{\textwidth}{Xccccc}
\toprule
\textbf{Covariate} & \textbf{Coefficient} & \textbf{Standard Error} & \textbf{Lower CI} & \textbf{Upper CI} &\textbf{p-value} \\ 
\midrule
IRD                                      & 0.00773                                  & 0.00280   & 0.00225  & 0.01321  & 0.00571                             \\ 
ER\_Regime                             & 0.11504                                 & 0.01754   & 0.08066  & 0.14942  & 0.00000                            \\ 
Inf\_diff                      & 0.00023                                  & 0.00151  & -0.00274 & 0.00320                                                                         & 0.88093                             \\ 
\bottomrule
\end{tabularx}
\noindent{\footnotesize{Note: CI = 95\% confidence interval}}
\end{table}

These results are in contrast with \citep{Glick2005}, who view export proceeds as a form of restriction on international capital flows and are positively correlated with an exchange rate crisis. Another study that contrasts our results is \citep{Edwards2009} which shows that a tightening of capital controls increases the volatility of exchange rates. However, these findings are consistent with \citep{Panggabean2024}, which likewise report no evidence of the impact of repatriated export proceeds on exchange rate volatility in Indonesia in the short and immediate terms. 

\section{Robustness and sensitivity analysis}\label{robust}
\subsection{Diagnostics}
\begin{figure}[h!]
	\includegraphics[width=\textwidth]{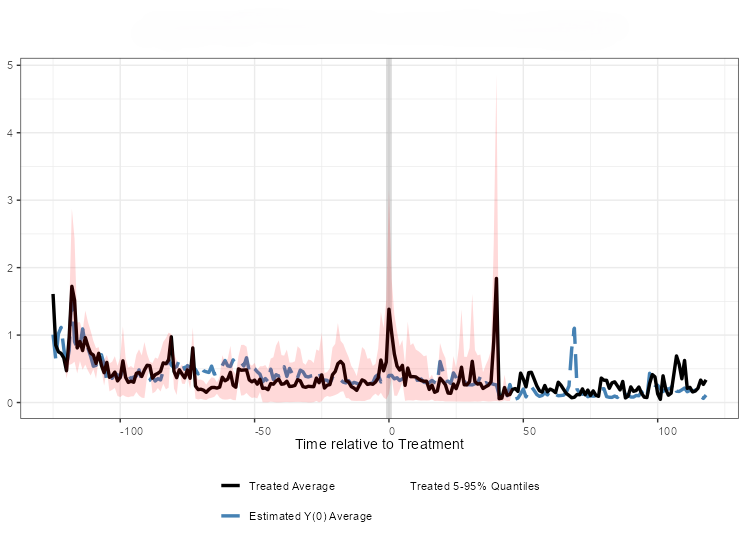}
	\caption{Treated and counterfactual averages with 95\% quantiles}
	\label{fig:counterfactual}
\end{figure}

By examining the counterfactual estimation in Figure~\ref{fig:counterfactual}, we can conclude that the estimated counterfactuals reproduce the treated trajectory reasonably well, where exceptions are noted only for a handful of occasions. Figure~\ref{fig:att} and Figure~\ref{fig:individual} can also be seen as a plot for dynamic treatment effects as defined by \citep{Liu2024}. From the figures, we observe that pretreatment residuals oscillate tightly around the zero line and show no discernible upward or downward drift; the figures provide little evidence of anticipatory behaviour or misspecified pre‑trends. The absence of systematic deviation confirms the strict exogeneity assumption. 

\begin{figure}[h!]
	\includegraphics[width=\textwidth]{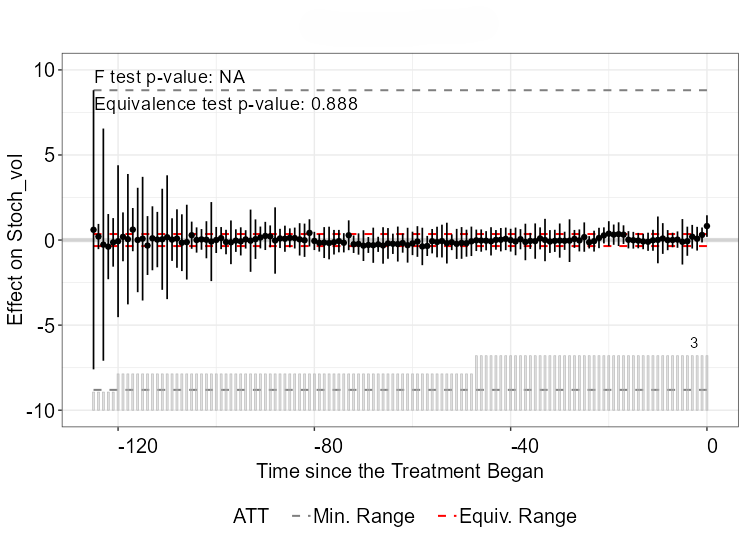}
	\caption{Equivalence Test}
	\label{fig:equivalence}
\end{figure}

Figure~\ref{fig:equivalence} reports an equivalence test whose high p-value confirms that the ATT is not statistically different from zero. However, statistical insignificance alone does not guarantee that the effect is practically irrelevant. The 95 \% confidence bands repeatedly cross the pre-specified equivalence margins (red dotted lines); therefore, we cannot show that all ATTs fall inside the indifference zone. In summary, the test leaves us with two unresolved possibilities: the policy may exert a non-zero influence, or any influence may be too small to matter substantively. The overall evidence is therefore inconclusive. 

\begin{figure}[h!]
	\includegraphics[width=\textwidth]{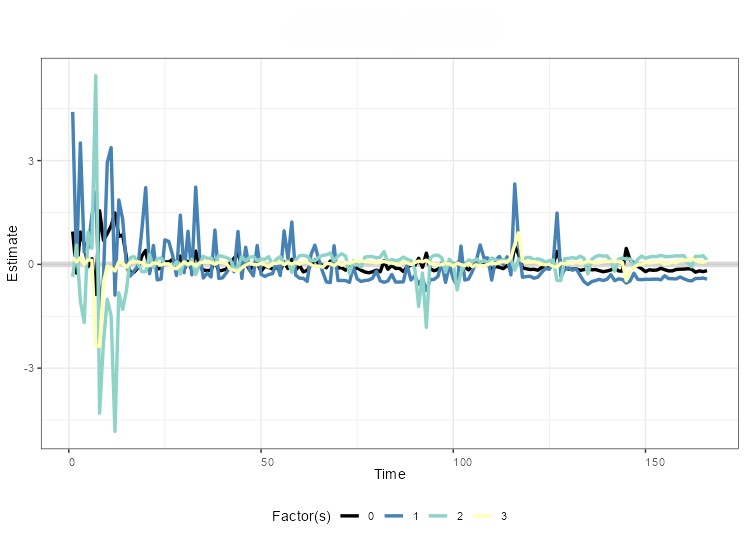}
	\caption{Latent Factors}
	\label{fig:latent}
\end{figure}

\begin{figure}[h!]
	\includegraphics[width=\textwidth]{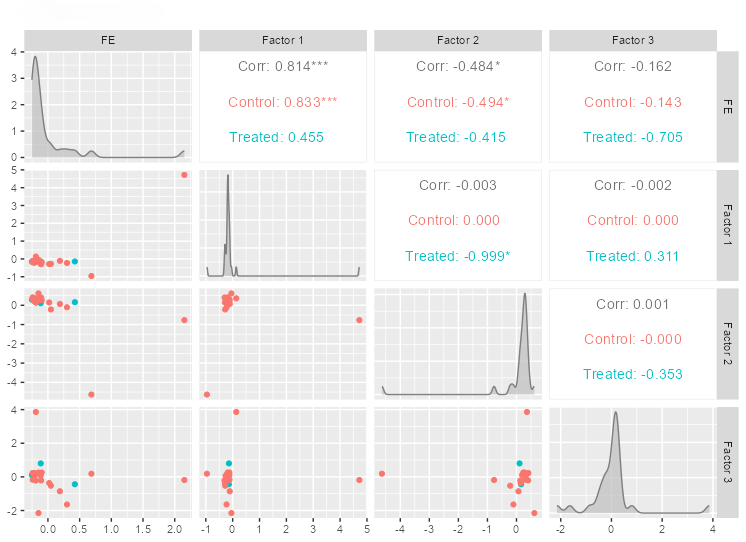}
	\caption{Factor loadings}
	\label{fig:loading}
\end{figure}

The three common factors we obtained in Subsection~\ref{result} influence exchange rate volatility and EPR policies, while unobserved and can capture unit-specific sensitivities. Figure~\ref{fig:latent} and Figure~\ref{fig:loading} show that the first factor exhibits pronounced early-sample volatility, which correlates strongly with the unit fixed effects, suggesting a global volatility shock that dissipates after the mid-sample. The second factor generates relatively greater volatility compared to the first factor at the beginning of the sample period and moderate, intermittent peaks around the mid-sample period. Moreover, the third factor captures only a one-off negative dip in the early period and minor fluctuations thereafter. After $t \approx 50$ all factors hover close to zero, indicating that most systematic comovement is confined to the initial turbulence.

These heterogeneous loadings underscore the necessity of integrating the IFE into our GFC model.  Cross-sectionally, the three factors display distinct loading distributions. Most treated loadings lie inside the control cloud, implying that the donor pool (control units) spans the treated units’ latent characteristics. The strongest relationship is between the unit fixed effect and Factor 1 ($\rho \approx 0.81$ for the whole sample): countries with higher average volatility also load more heavily on the global shock embedded in Factor 1. The correlations with Factors 2 and 3 are weak, which is consistent with these components capturing idiosyncratic, short-lived disturbances. Crucially, treated-unit loadings do not cluster at the extremes of any factor; hence, there is no evidence that the treated group is an outlier in the latent space. Therefore, the GSC should provide credible counterfactuals. In summary, the factor paths and loadings confirm that the model has isolated a small set of common shocks (mainly in the early sample) and that the treated economies are well represented within that latent-factor span.

\subsection{Sensitivity analysis}
Since the GSC works well even with a single treated unit, we employ the method for each treated unit separately but with the same covariates and the same control units. We then examine the results presented in Table~\ref{sensitive}. For Iran, Sri Lanka, and Turkey, there are very high p-values, indicating that there are no effect of EPR policies on exchange rate volatility, which is consistent with our results in Subsection~\ref{result}.  The MSPE remains low and the number of factors suggested by the cross validation procedures (3, 3, and 4), are similar to the number of factors that we used in our model. The significance of the coefficients of the covariates is also consistent with our results. All three countries show that the inflation differential is not statistically significant different from zero as a predictor of exchange rate volatility. 

\begin{table}[h!]
\caption{Overview of GSC models implemented on each treated unit}
\begin{adjustwidth}{-\extralength}{0cm}
 \label{sensitive}
\begin{center}
\begin{tabularx}{\fulllength}{*{4}{>{\centering\arraybackslash}c}}
 \toprule
\textbf{Model}  & \multicolumn{3}{c}{\textbf{Response Variable: Exchange Rate Volatility}}\\                                                                                                                                                                                                                                                         
\cmidrule{2-4} 
  & \textbf{Iran}                                   &\textbf{Sri Lanka}                                                                         & \textbf{Turkey}\\                                              
 \midrule
 ATT                                 &-0.02547 & 0.03486   &0.27393\\
					&(se: 0.17905) 			&(se: 0.19133) 				& (se: 0.35411)\\
					& [p-value: 0.88686] 		&[p-value: 0.85544] 			& [p-value: 0.43917]\\
 MSPE                                & 0.04919                            & 0.02911                          & 0.12641                                                                                  \\  
sigma2                              & 0.08534                           &0.10236                         & 0.10236                                                                                     \\  
Covariates                          & \multicolumn{3}{c}{IRD, ER\_Regime, Inf\_diff}                                                                                                                                                                                                                                                                                      \\
Unobserved factors           &4                                       & 3                                         & 3                                                                                           \\ 
 Significant $\beta$          & IRD, ER\_Regime           & IRD, ER\_Regime              & IRD, ER\_Regime                                                                              \\ 
Fixed effects                       & \multicolumn{3}{c}{two way}                                                                                                                                                                                                                                                                                                       \\
Observations                        & \multicolumn{3}{c}{166}                                                                                                                                                                                                                                                                                                           \\ 
Treated country                     & Iran                              & Sri Lanka                & Turkey                                                                                      \\  
 Control countries                  & \multicolumn{3}{X}{Armenia, Bahrain, Bolivia, Botswana, Gambia, Georgia, Jamaica, Jordan, Kenya, Kuwait, Kyrgyzstan, Mexico, Mongolia, Oman, Paraguay, Peru, Philippines, Qatar, Saudi Arabia, Trinidad and Tobago, Uganda, United Arab Emirates, Uruguay, Zambia}                     \\
\bottomrule 
\end{tabularx}
\end{center}
\end{adjustwidth}
\noindent{\footnotesize{Note: sigma2 is the estimated variance of the error term}}
\end{table}

We also perform a sensitivity analysis of our model using an alternative covariate specification that includes IRD, prevailing exchange rate regime, and terms of trade (ToT). Since Iran, Sri Lanka, and Turkey do not rely heavily on exports, including ToT as a covariate is a way to disturb the model and check the reliability of our model. Our results using this set of covariates show that the ATT remains not significantly different from zero, whereas the same set of covariates, IRD and exchange rate regime, remain significant in determining exchange rate volatility. A summary of the results is presented in Table~\ref{sensitive2}

\begin{table}[h!]
\caption{Estimated average treatment effect on the treated with alternative covariates}
 \label{sensitive2}
\begin{tabularx}{\textwidth}{ccccc}
\toprule
\textbf{ATT} & \textbf{Standard Error} & \textbf{Lower CI} & \textbf{Upper CI} & \textbf{p-value} \\ 
\midrule
0.08713                    & 0.07644                          & -0.06269              & 0.23695                & 0.25435       \\ 
\bottomrule
\end{tabularx}
\noindent{\footnotesize{Note: CI = 95\% confidence interval}}
\end{table}

\subsection{Placebo Tests}
To verify whether our GSC estimates are not driven by spurious fit, we perform both in-time and in-space placebo tests. Table~\ref{placebo1} presents the results of the in-time placebo test when we pretend that the policy began for every treated country in January 2013, which is twelve periods before the earliest actual adoption. The placebo average treatment effect is small and its bootstrap p-value far exceeds the 95\% significance thresholds, indicating that the GSC procedure does not generate false positives when no treatment has yet occurred.

\begin{table}[h!]
\caption{Estimated average treatment effects on the treated for in-time placebo test}
 \label{placebo1}
\begin{tabularx}{\textwidth}{ccccc}
\toprule
\textbf{Placebo ATT} & \textbf{Standard Error} &\textbf{Lower CI} &\textbf{Upper CI} &\textbf{Placebo p-value} \\ 
\midrule
0.05805         & 0.17686                          & -0.28859             & 0.40469               & 0.74275                             \\ 
\bottomrule
\end{tabularx}
\end{table}

Following \citep{Abadie2015}, we perform in-space placebo tests in which we assign the policy, one at a time, to each control country, using the three actual start dates of Iran, Sri Lanka, and Turkey. For every such pseudo-treated unit, we re-estimate the model and record the placebo ATT. The empirical p-value is computed as the share of placebo ATTs whose value is at least as large as the true ATT. We discard placebo runs for which cross-validation selects r = 0 factors (i.e., ordinary two-way fixed effects) because in those cases the GSC specification is not applicable. Table~\ref{placebo2} shows the placebo effects for each pseudo-treated unit, and that the empirical p-value is not statistically significant.

\begin{table}[h!]
\caption{Details of estimated average treatment effects on the treated for in-space placebo tests}
\begin{adjustwidth}{-\extralength}{0cm}
 \label{placebo2}
\small
\begin{tabularx}{\fulllength}{*{7}{>{\centering\arraybackslash}X}}
\cmidrule{1-7}
\textbf{Pseudo-treated Unit}                           & \multicolumn{2}{X}{\textbf{First Treatment Time: Apr 2018}} & \multicolumn{2}{X}{\textbf{First Treatment Time: Jan 2014}} & \multicolumn{2}{X}{\textbf{First Treatment Time: Sep 2018}} \\ 
\cmidrule{2-7}
 &\textbf{Placebo ATT}                & \textbf{p-value*}                & \textbf{Placebo ATT}                 & \textbf{p-value*}                & \textbf{Placebo ATT}                & \textbf{p-value*}                            \\ 
\cmidrule{1-7}
Armenia                    & -0.00056                    &1                                & -0.03748                     & 1                 & -0.05940                    & 1                                            \\ 
Bahrain                     &-0.00688                   &1                                & -0.00960                     & 1                     &-0.01305                   & 1                                            \\ 
Bolivia                        & -0.00667                    & 1                                & 0.00113             & 1                           & -0.01251                  & 1                                            \\ 
Botswana                   & -0.15773                    & 1                                & -0.23320            & 1                        & -0.16393                      & 1                                            \\ 
Gambia                     & -2.25546                    & 0                                & -3.28953            & NA                        & -2.32469                      & 0                                            \\ 
Georgia                    & 0.02534                     & 1                                & 0.25366               & 1                         & 0.02604                         & 1                                            \\ 
Jamaica              & 0.27812                     & 1                                & -0.01902                     & 1                         & 0.27374                       & 1                                            \\ 
Jordan                    & 0.00005                     & 1                                & 0.00534                      & 1                    & -0.00350                         & 1                                            \\ 
Kenya                     & -0.13958                    & 1                                & -0.14343                     & 1                   & -0.15849                          & 1                                            \\ 
Kuwait                     & -0.13324                    & 1                                & -0.08446                    & 1              & -0.14757                                 & 1                                            \\ 
Kyrgyzstan               & 0.24428                    & 1                               & 0.12806                      & 1            & 0.25396                                 & 1                                            \\ 
Mexico                       & 0.07573                    & 1                               & -0.12681                    & 1               & 0.05667                                  & 1                                            \\ 
Mongolia                  & -0.00852                    & 1                                & -0.03972                     & 1              & 0.04329                                  & 1                                            \\ 
Oman                         & -0.00091                   & 1                                & -0.00175                    & 1                & -0.00271                                & 1                                            \\ 
Paraguay                   & -0.15717                   & 1                              & -0.25431                    & 1                  & -0.14856                              & 1                                            \\ 
Peru                           & 0.10523                    & 1                               & 0.09762                      & 1                & 0.12142                                 & 1                                            \\ 
Philippines               & 0.00569                    & 1                                & -0.01584                    & 1                & 0.01251                                  & 1                                            \\ 
Qatar                        & 0.02965                     & NA                               & 0.11598                      & 1                & 0.04394                              & NA                                           \\ 
Saudi Arabia                   & -0.01111                   & 1                               & -0.00547                    & 1                & -0.01541 & 1                                            \\ 
Trinidad and Tobago         & -0.06644                    & 1                                & 0.06387                     & 1                                & -0.07147                                 & 1                                            \\ 
Uganda                         & -0.11615                  & 1                             & -0.31584                    & 1                       & -0.11658 & 1                                            \\ 
United Arab Emirates           & -0.00261           & 1                                & -0.00051                     & 1                       & -0.00530 & 1                                            \\ 
Uruguay                           & -0.44698                  & 0                               & -0.37769                     & 1                      & -0.45942 & 0                                            \\ 
Zambia                         & -0.30535                  & 1                              & 0.53493                     & NA                               & -0.28824 & 1                                            \\ 
\multicolumn{7}{c}{ATT of in-sample placebo test: -0.10711}                                                                                                                                                                                                                                                                                                                                                                \\ 
\multicolumn{7}{c}{p-value of in-sample placebo test: 0.94118}                                                                                                                                                                                                                                                                                                                                                             \\ 
\cmidrule{1-7}
\end{tabularx}
\end{adjustwidth}
\noindent{\footnotesize{Note: p-value* is 1 if placebo ATT $\geqslant$ true ATT and 0 otherwise; p-value* is NA when $r=0$, indicating the infeasibility of GSC implementation for such pseudo-treated unit.}}
\end{table}

Both placebo tests strengthen confidence in our baseline findings. That is, the non-existence of a robust impact of EPR policies on exchange-rate volatility is not caused by model overfitting or by chance correlations in the data.

\section{Conclusions}
This study investigates the impact of an export proceeds repatriation policy on exchange rate volatility in Iran, Sri Lanka, and Turkey from March 2008 to December 2021. Our results do support the hypothesis that the export proceeds repatriation policy does not have an impact on exchange rate volatility. However, we cannot reject the possibility that any impact is too small to be substantively meaningful. These conclusions are robust across several sensitivity analyses and placebo tests, attesting to their reliability. These results are in contrast with several studies on foreign capital flows that flag EPR policies as a form of restriction on international capital flows that increases exchange rate volatility. However, another former study on repatriated export proceeds shows similar results to ours.

Nevertheless, our results should be interpreted with caution. The absence of proprietary export proceeds volume data and central bank intervention may influence the results, given the different natures of regulations in each country.  We suggest that further research include these data to gain deeper insights into each emerging country being analyzed.  
\appendix
\section{Data Source for Short-term Interest Rates}
\begin{table}[h!]
\caption{Data source for short-term interest rates}
 \label{ird}
\begin{adjustwidth}{-\extralength}{0cm}

\begin{tabularx}{\fulllength}{cccc}
\toprule

\textbf{No.} 	&\textbf{Country} 	&\textbf{Interest Rates} 	&\textbf{Data Sources}		\\ 
\midrule
1            		& Armenia                                    		& 	Money market rates				& CEIC								\\ 
2            		& Bahrain                                    		& 	 BHIBOR						& Refinitiv Eikon					 		\\ 
3            		& Bolivia                                   		& 	Interbank rates					& CEIC								 \\ 
4            		& Botswana                                 		& 	Policy rates						& CEIC				 				\\ 
5            		& Gambia                                     		& 	T-bill rates and Policy rates			& CEIC								 \\ 
6            		& Georgia                                   		& 	Money market rates				& Bloomberg							 \\ 
7				& Iran										&  	Policy rates, Interbank rates, and Deposit rates		&	 CEIC and central bank website							\\ 
8            		& Jamaica                                    		& 	Money market rates and T-bill rates		& CEIC								 \\ 
9            		& Jordan                                    		& 	Money market rates				& CEIC								 \\ 
10            		& Kenya                                  			& 	Interbank rates					& CEIC								 \\ 
11            		& Kuwait                                    		& 	Interbank rates					& CEIC								 \\ 
12           		& Kyrgyzstan                                   		& 	Interbank rates					& CEIC								 \\ 
13            		& Mexico                                    		& 	Money market rates and TIIE			& CEIC and Refinitiv Eikon					 \\ 
14            		& Mongolia                                    		& 	 Interbank rates 					& CEIC								 \\ 
15            		& Oman                                   		& 	Interbank rates					& CEIC								 \\ 
16            		& Paraguay                                     		& 	Money market rates and Policy rates		& CEIC								 \\ 
17            		& Peru                                    			& 	Interbank rates					& Refinitiv Eikon							 \\ 
18            		& Philippines                                    		& 	Money market rates				& CEIC						 		\\ 
19            		& Qatar                                    		& 	Overnight Deposit rates and QIBOR		& Refinitiv Eikon							 \\ 
20            		& Saudi Arabia                                   	& 	Policy rates, Money market rates, and SAIBOR & CEIC and Refinitiv Eikon					 \\ 
21				& Sri Lanka							&	SLIBOR and Call market rates				&	CEIC							\\ 
22            		& Trinidad and Tobago                              	& 	Reverse Repo rates				& Bloomberg							 \\ 
23				& Turkey							&		Interbank rates and Money market rates	& CEIC									 \\ 
24            		& Uganda                                   		& 	T-bill rates and Overnight Deposit rates	& CEIC and Refinitv Eikon					 \\ 
25            		& United Arab Emirates                            	& 	Interbank rates and EIBOR			& CEIC and central bank website				\\ 
26            		& Uruguay                                    		& 	Money market rates and Interbank rates	& CEIC and Refinitv Eikon					 \\ 
27            		& Zambia                                   		& 	Policy rates and Overnight Deposit rates	& CEIC and Refinitv Eikon					 \\ 
\bottomrule
\end{tabularx}
\end{adjustwidth}

\noindent{\footnotesize{Note: BHIBOR=Bahraini Interbank Offered Rate; TIIE=Tasa de Interes Interbancaria de Equilibrio (Interbank Equilibrium Interest Rate); QIBOR=Qatar Interbank Offered Rate; SAIBOR=Saudi Arabian Interbank Offered Rate; SLIBOR= Sri Lankan Interbank Offered Rate; EIBOR=The Emirates Interbank Offered Rate.}}
\end{table} 

\clearpage
\bibliographystyle{elsarticle-harv} 
\bibliography{dhe}

\end{document}